\documentstyle[aps,prl,preprint,floats,epsfig]{revtex}

\begin{document}

%\draft

% Comment out to get double spacing.
\tighten

\title{Measurements of the Mass, Total Width and Two-photon Partial \\ 
Width of the $\eta_c$ Meson}  

% Insert author list file here - PRL style author list
\author{
G.~Brandenburg,$^{1}$ A.~Ershov,$^{1}$ Y.~S.~Gao,$^{1}$
D.~Y.-J.~Kim,$^{1}$ R.~Wilson,$^{1}$
T.~E.~Browder,$^{2}$ Y.~Li,$^{2}$ J.~L.~Rodriguez,$^{2}$
H.~Yamamoto,$^{2}$
T.~Bergfeld,$^{3}$
% ...
}
 
\author{(CLEO Collaboration)}

\address{
$^{1}${Harvard University, Cambridge, Massachusetts 02138}\\
$^{2}${University of Hawaii at Manoa, Honolulu, Hawaii 96822}\\
$^{3}${University of Illinois, Urbana-Champaign, Illinois 61801}\\
% ...
}
 
\date{\today}
\maketitle

\begin{abstract} 
% Insert abstract here.
Using 13.4 $\rm fb^{-1}$ of data collected with the CLEO detector at 
the Cornell Electron Storage Ring, 
we have observed 300 events for the two-photon production of ground-state 
pseudo-scalar charmonium in the decay 
$\eta_c \to K_{S}^0K^{\mp}\pi^{\pm}$. 
We have measured the $\eta_c$ mass to be 
(2980.4 $\pm$ 2.3 (stat) $\pm$ 0.6 (sys)) MeV 
and its full width as 
(27.0 $\pm$ 5.8 (stat) $\pm$ 1.4 (sys)) MeV. 
We have determined the two-photon partial width of the $\eta_c$ meson to be 
(7.6 $\pm$ 0.8 (stat) $\pm$ 0.4 (sys) $\pm$ 2.3 (br)) $\rm keV$,
with the last uncertainty associated with the decay branching fraction.

\end{abstract}


% Begin main body of text.
In this Letter, we report a study of two-photon production of the 
ground-state pseudo-scalar charmonium, 
{\it i.e.,} $\gamma\gamma \rightarrow \eta_c$. The two space-like 
photons are radiated by $e^+$ and $e^-$ beams,
each at an energy of approximately 5.3 GeV.
The charmonium 
spectrum is an ideal testing ground for 
quantum chromodynamics (QCD) calculations, and 
producing C-even charmonium states through $\gamma\gamma$ fusion 
provides a clean environment for this purpose.

The two-photon partial width of the $\eta_c$ meson
can be expressed in next-to-leading order (NLO) perturbative QCD
(PQCD), in terms of the $e^+e^-$ partial width of 
the $J/\psi$ meson, as\cite{Kwong} 
\begin{equation}
\frac{\Gamma_{\gamma\gamma}^{\eta_c}}{\Gamma_{ee}^{\psi}} 
= \frac{4}{3}(1 + 1.96\alpha_s/\pi)
\times\frac{|\Psi_{\eta_c}(0)|^2}{|\Psi_{\psi}(0)|^2}.
\end{equation} 
Using the world average value\cite{PDG} of $\Gamma_{ee}^{\psi}$, 
a value of 
the strong coupling constant
$\alpha_s$ evaluated at the charm mass scale\cite{Kwong} of 
(0.28 $\pm$ 0.02), and the assumption that the two 1S 
wave functions, 
$\Psi$, 
are the same at the origin, 
this relationship predicts 
$\Gamma_{\gamma\gamma}^{\eta_c} = (8.2 \pm 0.6)$ keV. 
%Removed reference to relativistic corrections.
%Relativistic corrections 
%%tend to 
%reduce the $\Gamma_{\gamma\gamma}^{\eta_c}$ 
%estimation slightly. 

The total width of the $\eta_c$ meson
can be assumed to be dominated by its two-gluon component, {\it i.e.}, 
$\Gamma_{\rm tot}^{\eta_c} \approx \Gamma_{gg}^{\eta_c}$. 
The ratio 
of the rates 
for $\eta_c \to gg$ and $\eta_{c} \to \gamma\gamma$ 
is an especially clean prediction of PQCD
because the dependencies of these rates on the wave functions
and non-perturbative factors are identical in the
numerator and denominator.
The ratio depends only on the coupling constants and
has been calculated in NLO\cite{Kwong}, 
\begin{equation}
\frac{\Gamma_{\rm tot}^{\eta_c}}{\Gamma_{\gamma\gamma}^{\eta_c}} \approx 
\frac{9 \alpha_s^2}{8 \alpha^2} \times \frac{(1 + 4.8 \alpha_s/\pi)}{(1-3.4\alpha_s/\pi)}.
\end{equation} 
Using the value of $\Gamma_{\gamma\gamma}^{\eta_c}$ estimated in NLO 
gives $\Gamma_{\rm tot}^{\eta_c}$ as (28 $\pm$ 6) MeV;
using instead the world average value\cite{PDG} of 
$\Gamma_{\gamma\gamma}^{\eta_c}$, 
one obtains an estimate of 
$\Gamma_{\rm tot}^{\eta_c}$ as (26 $\pm$ 6) MeV. 
A calculation 
%done 
with fully relativistic decay amplitudes
and a sophisticated QCD potential model\cite{Gupta} predicts 
$\Gamma_{\rm tot}^{\eta_c} \approx 23$ MeV.  The 
current world average\cite{PDG} of 
$\Gamma_{\rm tot}^{\eta_c} = 13.2_{-3.2}^{+3.8}$ MeV  disagrees 
with these theoretical expectations. A precise measurement of the 
full width and two-photon partial width of the $\eta_c$ is
important for the verification of these 
PQCD calculations and approximations.

% Order of these next two paragraphs changed
In the two-photon process 
$e^+e^-$ $\rightarrow$ $e^+e^-\gamma\gamma$ $\rightarrow$ $e^+e^-\eta_c$, 
the photon propagators dictate that
the cross section naturally peaks at low momentum transfer,
so the photons are almost real (``on shell''). 
The incident leptons are scattered at very low angles and
continue traveling down the beam pipe undetected. Such ``untagged'' events 
typically have low net transverse momentum and low visible energy. 
The production of the $\eta_c$ meson 
in this untagged two-photon process 
was searched for in the $K_{S}^0K^{\mp}\pi^{\pm}$ decay mode.

The data used in this study correspond to an integrated luminosity of 
13.4 $\rm fb^{-1}$ and were collected with two configurations 
(CLEO II\cite{CLEOII} and CLEO II.V\cite{CLEOII.V}) of the CLEO 
detector at the Cornell Electron Storage Ring (CESR). Approximately 
one third of the data were taken with the CLEO II configuration.  The 
detector components most useful for this study were 
the concentric tracking devices for charged particles, operating in
a 1.5T superconducting solenoid.
For CLEO II, 
this tracking system consisted of a 6-layer straw tube chamber, a 10-layer 
precision drift chamber, and a 51-layer main drift chamber. The main 
drift chamber also provided measurements of the specific ionization 
loss, $dE/dx$, used for particle identification. For CLEO II.V, the straw 
tube chamber was replaced by a 3-layer, double-sided silicon vertex 
detector and the gas in the main drift chamber was changed from a 
50:50 mixture of argon-ethane to a 60:40 helium-propane mixture. These
changes gave rise 
to significant improvements in the momentum and $dE/dx$ resolutions 
for charged tracks. 
Photons were detected using the high-resolution electromagnetic 
calorimeter consisting of 7800 CsI crystals. The Monte Carlo simulation 
of the CLEO detector response was based upon GEANT\cite{GEANT}. 
Simulated events were processed in the same manner as the data 
to determine 
the $\eta_{c} \to K_{S}^{0}K^{\mp}\pi^{\pm}$ detection efficiency
and the $K_{S}^{0}K^{\mp}\pi^{\pm}$ mass resolution at the
$\eta_{c}$ mass.

The $K_{S}^0$ vertex was reconstructed from its decay to $\pi^+\pi^-$ and 
was required to be displaced from the $e^+e^-$ interaction point; the
amount of this displacement varied with detector configuration but
was $\approx$1.5 mm.
The 
$K_{S}^0$ candidate was also required to be within 4 standard deviations 
($\sigma$) of the known $K_{S}^0$ mass\cite{PDG};
here $\sigma$ was determined on an event-by-event basis
from the momenta measurements.
Furthermore, the $K_{S}^0$ momentum 
vector was required to point back to the interaction point. 
Of the two remaining charged tracks,
the $K^{\mp}$ and $\pi^{\pm}$ candidates,
the one with lower momentum was 
typically uniquely identified using the 
particle's specific ionization ($dE/dx$). This fixed the identity of 
the only remaining unidentified track, because the 
presence of the 
$K_{S}^0$ dictated that exactly one of these two be a kaon to conserve 
strangeness in the event. 

The background from processes other than two-photon
production was suppressed by requiring 
that the $\eta_c$ 
candidate have net transverse momentum less than 0.6 GeV/$c$ 
and that visible 
energy in the event be less than 6 GeV. 
Also, because the final state had 
no expected energy deposits in the calorimeter from
neutral particles, 
the total calorimeter energy in the event not matched to
charged tracks was required 
to be less than 0.6 GeV. 

\begin{figure}[ht] 
  \centerline{\epsfig{file=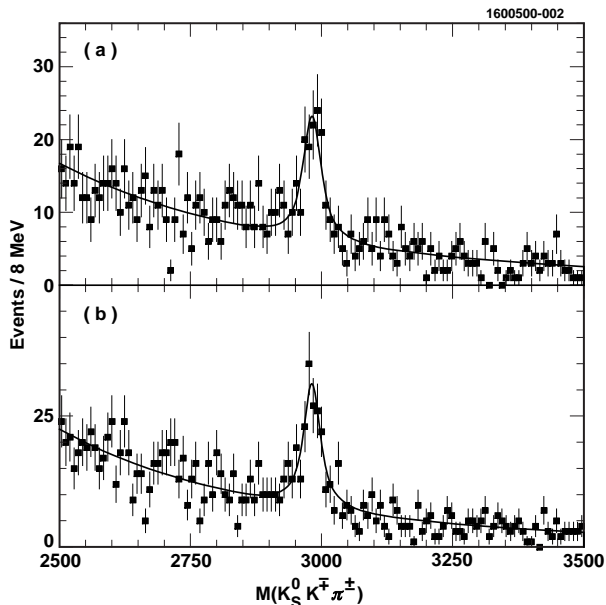,height=3.2in}}
  \caption{Results of the simultaneous fit to $\eta_c$ candidates in 
(a) CLEO II and (b) CLEO II.V for width and yield measurements, 
with a combined $\chi^2$/d.o.f. = 226/243.}
  \label{mass}
\end{figure}

For the mass measurement only, we restricted 
ourselves to those events in which the 
$K^{\mp}$ and $\pi^{\pm}$
daughters of the $\eta_c$ candidate
traversed all layers of the tracking volume. 
The $K_{S}^0$ daughter pions were not required to satisfy the same 
criterion, in that the kinematic fitting of the $K_{S}^0$ 
decay corrected 
for any possible momentum mis-measurement of the daughter tracks. 
We did not make any such requirements while measuring 
%the width and two-photon partial width, 
$\Gamma_{\rm tot}^{\eta_c}$ and $\Gamma_{\gamma\gamma}^{\eta_c}$,
because these quantities 
are relatively insensitive 
to precise measurements of the track momenta; 
the distribution of candidate invariant masses for the
determination of these two quantities is shown in Fig. \ref{mass}.

We fitted the background with a power law function 
($A \cdot W_{\gamma\gamma}^n$, with $W_{\gamma\gamma}$ 
the $K_{S}^0K^{\mp}\pi^{\pm}$ invariant mass and $A$ a 
multiplicative constant) and the signal 
with a spin-0 relativistic 
Breit-Wigner function (describing the natural line shape) convolved 
with a double Gaussian function (to 
take into account the detector resolution). 
The parameters for this double Gaussian were obtained from 
a Monte Carlo sample that had 
the $\eta_c$ intrinsic width set to zero. 
We performed a simultaneous, binned, 
maximum-likelihood fit to the invariant mass distributions for the CLEO II 
and CLEO II.V datasets, constraining the physical variables 
$M_{\eta_c}$, 
$\Gamma_{\rm tot}^{\eta_c}$, and 
$\Gamma_{\gamma\gamma}^{\eta_c}$ 
in the two datasets to be the same. 
The constraint on $\Gamma_{\gamma\gamma}^{\eta_c}$
was accomplished by requiring the ratio of the fitted yields to be 
the same as the ratio of the integrated
luminosities of the two data sets times the efficiencies 
as determined from 
our simulations.
The invariant mass resolution 
was approximately 
9 MeV in CLEO II and 7 MeV in CLEO II.V. The bin width for fitting was 
chosen as approximately the average of these two resolutions. 

As noted above, two separate sets of fits were performed, one for the 
measurements of 
$\Gamma_{\rm tot}^{\eta_c}$ and $\Gamma_{\gamma\gamma}^{\eta_c}$
and another for the determination of the $\eta_{c}$ mass.  
The full width and yield were 
obtained from the distributions shown in Fig. \ref{mass}, with the 
total
observed 
yield being $N_{obs} = 300 \pm 32$.  The fit to the width gives 
$\Gamma_{\rm tot}^{\eta_c} = (27.0 \pm 5.8)$ MeV, with the uncertainty
being only statistical.  The two-photon partial width was determined 
by first correcting for the detector efficiency, $\epsilon$,
and then dividing by the number
of events expected, $N_1$,  for a two-photon partial width of 1 keV:
\begin{equation}
\Gamma_{\gamma\gamma}^{\eta_c} = N_{obs}/(\epsilon \cdot N_{1}).
\end{equation}
The quantity $N_{1}$ was determined using
\begin{equation}
N_{1} = {\mathcal{L}} \cdot {\cal B}_{\eta_c} \cdot {\cal B}_{K_{S}} 
\cdot \sigma_{e^{+}e^{-} \to e^{+}e^{-}\eta_{c}};
\end{equation}
${\mathcal{L}}$ is the integrated luminosity. 
The 
cross section for 
$e^{+}e^{-} \to e^{+}e^{-}\eta_c$ was obtained from Monte Carlo
simulation, using the formalism of
Budnev {\it et al.}\cite{BGMS} 
and setting $\Gamma_{\gamma\gamma}^{\eta_c} = 1$ keV; this
choice of this value has no effect on the extracted result.
Also, 
${\cal B}_{\eta_c} \equiv {\cal B}(\eta_c \to K_{s}^{0} K^{\mp} \pi^{\pm}) $
and 
${\cal B}_{K_{S}} \equiv {\cal B}(K_{S}^0 \to \pi^{+} \pi^{-})$,
with the world average values\cite{PDG} used. 
This procedure gives $\Gamma_{\gamma\gamma}^{\eta_c} = (7.6 \pm 0.8)$ keV,
with this uncertainty coming from statistics only.

The mass was obtained from fits to the more restrictive 
set of events, as described above, with a 
total signal size of 195 $\pm$ 24,
yielding  $M_{\eta_c} = (2980.4 \pm 2.3)$ MeV,
the uncertainty being statistical only.

Possible sources of systematic uncertainty for the measured mass, full 
width, and two-photon partial width were studied. The results are summarized 
in Table I, in which the individual uncertainties are added in quadrature to 
obtain the total systematic uncertainty. 

The mass calibration of our detector was checked by measuring the masses of the well 
known $K_{S}^0$, $\phi$, $D$, and $J/\psi$ mesons using decay modes involving
only charged tracks. The measurements 
were found to be in good agreement with their respective world averages when we 
limited ourselves to events in which the charged tracks traversed 
all layers of the tracking volume.  
The fitted mass including events outside this more restrictive
detector volume is consistent with the value we quote, but with
substantially larger systematic uncertainties.
Our particle identification procedure did not 
introduce any significant systematic bias to the mass measurement. The measured mass was 
also 
insensitive to variations in the signal shape used 
to fit the $\eta_c$ resonance. 

The systematic uncertainties in the measurement of the full width were dominated by 
effects due to the mass resolution, the particle identification procedure, and 
the signal shape used in the fit. 
Our ability to 
predict the actual mass resolution was tested by studying 
the reconstructed $D$ mesons in Monte Carlo simulation and data; the agreement 
was found to be better than 0.1 MeV. 

The particle identification procedure was
unable to distinguish between a charged kaon and a charged pion if the track 
momentum was above 0.8 GeV/$c$, for which the expected ionization losses are nearly 
equal for the two species. This led 
to a broadening of the reconstructed resonance and was taken into account by the 
wider Gaussian of the double Gaussian resolution function. This effect was limited to 
less than 5$\%$ of the events.  
We estimated the possible uncertainty due to this
mis-assignment of particle species by completely removing the 
fraction of events having two possible $\eta_{c}$ candidates
and assigning the corresponding change in the measured width as 
the systematic error.

The accuracy of the fitting method in extracting the Breit-Wigner width of the 
resonance was checked by extracting the $\eta_c$ widths from sets of 
simulation events generated with different intrinsic widths. We varied the parameters 
of the signal shape used to fit the $\eta_c$ resonance within their uncertainties, 
derived from a comparison of the fit to $D$ meson decays in Monte Carlo
simulation and data, 
to estimate the effect on the measured width. The measured width from the more 
restrictive set of events used for mass measurement was within 0.2 MeV of the 
corresponding measurement using the full sample, and no significant correlation was 
found between the measured mass and full width.

There were several sources of uncertainty for the estimation of the efficiency, 
which in turn affected the measurement of $\Gamma_{\gamma\gamma}^{\eta_c}$. 
These 
were dominated by the uncertainties in the tracking and trigger efficiencies. The 
effect of a possible presence of resonant substructure 
($K^*\overline{K}$) in $\eta_c$ decay 
was studied and was found to give an insignificant variation in the 
detection efficiency. We estimated the uncertainty in the measured partial
width from this effect by 
considering the possibility that all the $\eta_c$ 
mesons decay through $K^*\overline{K}$.
Our initial investigation showed roughly
a third of the $\eta_c \to K_{s}^{0} K^{\mp} \pi^{\pm}$ events
proceed via $K^{*}$(1430) with no evidence for any 
$K^{*}$(892); a detailed analysis of this substructure
is beyond the scope of this Letter.

The selection requirements on total visible energy and 
unmatched energy clusters were shown by simulation to be
essentially 100\% efficient for our signal process and free
of systematic bias.  Possible bias from the 
transverse momentum requirement
was investigated by changing the nature of the form factor in the
simulation and shown to also be negligible. 

In our analysis, we investigated the possible effects of
interference between the $K_{S}^0 K^{\mp}\pi^{\pm}$ resonant and
non-resonant final states. 
From the distribution of net
transverse momentum for events in the sidebands
of the signal lineshape, we
estimated that 
one third of the background events were not
of the type $\gamma\gamma \to K_{S}^0 K^{\mp}\pi^{\pm}$; these 
included events with at least one missing particle ($\pi_0$, $\gamma$) as
well as events of the type $e^{+}e^{-} \to$ hadrons, $e^{+}e^{-} \to
\tau^{+}\tau^{-}$, and $\gamma\gamma \to \tau^{+}\tau^{-}$. Such events
cannot have interference with our signal $K_{S}^0
K^{\mp}\pi^{\pm}$ events. 
Study of the helicity angle distributions in the
$\eta_{c}$ rest frame indicated that the
sideband events predominantly have $J=2$ (or higher) while our signal has
$J=0$.  Due to the preferential production of states with natural
parity in two-photon un-tagged processes, a majority of the 
remaining background
events were expected to have
the natural spin-parity ($0^+$) compared to the unnatural spin-parity 
($0^-$) of the
signal events.  The acceptance of our detector
is symmetric in polar angle and uniform in azimuth,
making the interference between these states of opposite parity vanish.
We 
did not include
any possible effects due to interference on the measured mass,
full width and two-photon partial width. Further, the signal shape showed
no distortions and the goodness of fit to the hypothesis that ignored
interference was very good, as shown in Fig. \ref{mass}.

In summary, we have measured the mass, full width, and two-photon partial width 
of the $\eta_c$ produced in two-photon collisions. The mass measurement of 
(2980.4 $\pm$ 2.3 (stat) $\pm$ 0.6 (sys)) ${\rm MeV}$ compares well with the 
world average\cite{PDG} of (2979.8 $\pm$ 2.1) ${\rm MeV}$. The measured total 
width of (27.0 $\pm$ 5.8 (stat) $\pm$ 1.4 (sys)) $\rm MeV$ disagrees with the 
world average\cite{PDG} of 
($13.2_{-3.2}^{+3.8}$) MeV, which consists of measurements 
with large 
relative uncertainties (40$-$100$\%$). Our measured width is consistent with 
theoretical expectations\cite{Kwong,Gupta}. The measured two-photon partial width 
of the $\eta_c$ meson of (7.6 $\pm$ 0.8 (stat) $\pm$ 0.4 (sys)) keV
agrees well with 
the world average\cite{PDG} of $(7.5^{+1.6}_{-1.4})$ keV and theoretically expected 
values, and is a significant improvement 
%in terms of the uncertainty.  
in terms of experimental precision.
We use 
the world average\cite{PDG} of the 
$\eta_c \rightarrow K \overline{K} \pi$ branching fraction of $(5.5 \pm 1.7)\%$. 
The uncertainty in $\Gamma_{\gamma\gamma}^{\eta_c}$ due to the 
uncertainty in this branching fraction 
is $\pm$2.3 keV 
and is stated separately from the other contributions.
From the ratio 
of our measured full width and two-photon partial width, we have 
extracted $\alpha_s$ 
at the charm mass scale to be $0.285 \pm 0.025$,
for which we have added our sources of uncertainty in
quadrature. We have used the NLO calculation in Eq. 2 
to estimate $\alpha_s$, thus making the result 
dependent on renormalization scheme and scale;
we have not included such theoretical uncertainties
in our quoted value.

Our measurements of $\Gamma_{\rm tot}^{\eta_c}$ and 
$\Gamma_{\gamma\gamma}^{\eta_c}$ show that 
PQCD calculations are able to reliably predict 
the ratios 
of the decay rates 
%of the basic parameters 
of a heavy quarkonium system, 
%where non-perturbative and relativistic corrections are expected to be small. 
where non-perturbative effects cancel.

We gratefully acknowledge the effort of the CESR staff in providing us with excellent 
luminosity and running conditions. This work was supported by the National Science 
Foundation, the U.S. Department of Energy, the Research Corporation,
the Natural Sciences and Engineering Research Council of Canada, the A.P. Sloan Foundation, 
the Swiss National Science Foundation, the Texas Advanced Research Program, and the 
Alexander von Humboldt Stiftung.

\begin{center}
\begin{table}
\smallskip
\begin{tabular}{|l|c|c|c|}
Source of                          &  $M_{\eta_c}$ 
& $\Gamma_{\rm tot}^{\eta_c}$ & $\Gamma_{\gamma\gamma}^{\eta_c}$\\
systematic uncertainty & (MeV) & (MeV) & (keV)\\
\hline
Mass calibration		& 0.6 		& $<0.1$ 	& $<0.1$\\
Particle identification		&  $<0.1$ 	& 1.3 		& 0.1\\
Signal shape			&  $<0.1$ 	& 0.3 		& 0.1\\
Detector resolution		&  $<0.1$ 	& 0.3 		& $<0.1$ \\
Trigger				&  -  		&  -  		& 0.2\\
Tracking                        &  -  		&  -  		& 0.2\\
Resonant substructure	        &  $<0.1$  	&  $<0.1$  	& 0.2\\
Luminosity			&  -  		&  -  		& 0.1\\
$K_{S}^0$ selection		&  $<0.1$  	&  $<0.1$  	& 0.1\\
Event Selection			&  $<0.1$  	&  $<0.1$  	& $<0.1$ \\
\hline\hline
Total                           &  0.6		& 1.4 		& 0.4\\
\end{tabular}
\caption{Systematic uncertainties in the three measurements.  The overall
value is obtained by adding the individual contributions in quadrature.}
\end{table}
\end{center}

% End main body of text
\end{document}